\documentclass[prl,twocolumn,showpacs]{revtex4}
\usepackage{graphicx}
\usepackage{color}
\usepackage{bm}
\usepackage{longtable}
\usepackage{graphics}
\usepackage{amsmath}
\usepackage{bm}
\usepackage{amsfonts}
\usepackage{amssymb}

\begin{document}

\date{\today}
\title{Communicating with a wave packet using quantum
superarrival}
\author{Dipankar Home
\footnote{dhome@bosemain.boseinst.ac.in}}
\affiliation{CAPSS, Department of Physics, Bose Institute, Sector-V, Salt Lake, Kolkata
700 091, India}
\author{A. S. Majumdar
\footnote{archan@bose.res.in}}
\affiliation{S. N. Bose National Centre for Basic Sciences, Block JD, Sector III, Salt
Lake, Kolkata 700098, India}
\author{A. Matzkin
\footnote{alexandre.matzkin@u-cergy.fr}}
\affiliation{LPTM (CNRS Unit\'{e} 8089), Universit\'{e} de Cergy-Pontoise, 95302 Cergy-Pontoise cedex, France}

\begin{abstract}
An analytical treatment of a propagating
wave packet incident on a
transient
barrier reveals a counterintuitive quantum mechanical
effect in which, for a particular time interval, the time-varying transmission
probability {\it exceeds} (`superarrival') that for the free propagation of
the wave packet. It is found that the speed with which the information
about the barrier perturbation propagates across the wave packet can
exceed the group velocity of the wave packet. An interesting implication
of this effect regarding information transfer is analyzed by showing
one-to-one correspondence between the strength of the barrier and the
magnitude of `superarrival'.
\end{abstract}

\pacs{03.65.-w,03.65.Ta,03.67.Hk}
\maketitle

\paragraph{A. Introduction.\textemdash} A number of interesting
phenomena have been uncovered in recent years using the dynamics of
quantum wave packets. Among these, a class of novel quantum effects
as in the revival of wave packets \cite{robinett} and quantum transients
\cite{muga} are worth mentioning. In particular, for propagating wave packets,
appropriate changes in the boundary conditions for suitable potentials can
give rise to curious dynamical features \cite{chen09,muga09,super}. One such striking effect, which we
call `quantum superarrival'
is analytically demonstrated in this paper by
considering a Gaussian wave packet which is incident on a time-dependent potential barrier.
For the purpose of the analytical treatment given in
this paper, the form of this barrier is chosen such
that it corresponds to a transient parabolic barrier acting over a small time
interval during which the peak of the propagating wave packet crosses the
maximum of the parabolic barrier. In this case, we find that there exists an
interval of time during which there is an {\it enhancement}
(`superarrival') of the time-evolving transmission probability as compared to
the case of a wave packet freely propagating in the absence of any potential
barrier.

In the usual studies, the transmission/reflected probabilities for the
scattering of wave packets by potential barriers are calculated after a
complete time-evolution when the asymptotic values have been attained. In the
present work, based on the analytical solution of the relevant time-dependent
Schr{\" o}dinger equation, a phenomenon is displayed which occurs during the
time evolution of such a probability
that is found to have the following salient features. While the
effect of barrier perturbation resulting in `superarrival' is discernible by
measuring the transmission probability, it becomes more pronounced with the
increase of the rate at which the barrier perturbation occurs (in the case
considered, it is the strength of the barrier). Further, it is shown that
the effect of barrier perturbation propagates across the wave packet at a
speed that depends upon the strength of the barrier, thereby leading to a
new concept of what we call \textit{`information velocity'}.

In particular, for appropriate choices of the relevant parameters, it is shown
that this information velocity can be higher than the group velocity of the
incident wave packet, thereby illustrating that the information content of
a wave packet does not always propagate with the group velocity of a wave
packet. Here a local change in the potential affects a wave packet globally
through its time evolution where the wave function plays the role of a
\textit{carrier} through which the information about the barrier perturbation is
transmitted. Interestingly, by exploiting this feature of `superarrival', it is possible to
develop a scheme for communication whose basic idea is indicated in this paper.
For this, we proceed by first delineating the relevant details of the
analytical treatment that leads to the phenomenon of `superarrival' in the
example we consider in this paper.

\paragraph{B. Superarrival: the phenomenon.\textemdash}

We begin our analysis by considering a Gaussian wave packet peaked at $q_{0}$%
\begin{equation}
\psi(x,t_{0})=\left( \frac{2m}{\pi\alpha_{0}^{2}}\right) ^{1/4}e^{-m\left[
x-q_{0}\right] ^{2}/\alpha_{0}^{2}}e^{ip_{0}\left[ x-q_{0}\right] /\hslash }
\label{1}
\end{equation}
which is incident on  a time-dependent barrier given by
\begin{equation}
V(x,t)=-\frac{1}{2}mke^{-g(t-t_{b})^{2}}x^{2},  \label{30}
\end{equation}%
that corresponds to the appearance of a parabolic barrier during a
small time interval. This is achieved by choosing a Gaussian form for
the time window, with the parameters $t_{b}$ and $g$
indicating the peak time and inverse width of the window. $k$ determines
the barrier strength.

\begin{figure}[tb]
\includegraphics[height=9cm]{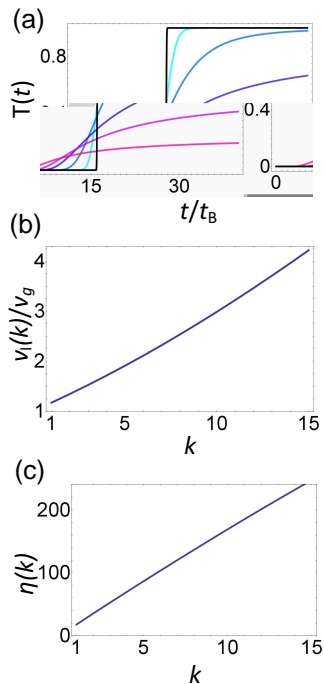}
\caption{Superarrivals for a system with parameters $q_0=-10^3, p_0=10, m=5\cdot10^{4}, \alpha^2(t_0)=10^7, t_B=5\cdot10^6, g=10^{-10}$ (atomic units).
(a): The transmission probability $T(t)$ is plotted for the free case
(black curve) and for several time-dependent barriers characterized by
different strengths: $k/10^{11}=1, 3, 6, 9, 15$ (the curve color goes from light blue
to red as $k$ increases). (b): The ratio $v_{I}(k)/v_{g}$ between the barrier perturbed
information velocity and the free group velocity is plotted as a function of the barrier strength $k$. (c): The magnitude
of superarrivals $\protect\eta$ is  plotted versus $k$.}
\label{figT}
\end{figure}

Having in mind the  path-integral
properties for quadratic Lagrangians (see below), the
time-dependent wavefunction is obtained from the ansatz
\begin{eqnarray}
\psi (x,t)=\left( \frac{2m}{\pi \alpha ^{2}(t)}\right) ^{1/4}e^{-\left[
x-q(t)\right] ^{2}\left( \frac{m}{\alpha (t)^{2}}-\frac{im\alpha ^{\prime
}(t)}{2\hslash \alpha (t)}\right) } \nonumber \\   \label{z0}
e^{ip(t)\left[ x-q(t)\right] /\hslash }
e^{\frac{i}{2\hslash }\left[ p(t)q(t)-p_{0}q_{0}\right] }e^{-i\left[ \phi
(t)-\phi _{0}\right] },
\end{eqnarray}%
It can be checked by
direct substitution that Eq (\ref{z0}) obeys the Schr\"{o}dinger equation with the initial condition
(\ref{1}), provided
$q(t)$ and $p(t)$ obey the classical equations of motion, i.e.,
\begin{equation}
\partial _{t}^{2}q(t)=\omega ^{2}(t)q(t)  \label{20}
\end{equation}%
and $\partial _{t}p(t)=m\partial _{t}^{2}q(t)$, with $q_{0}\equiv q(t_{0})$ and $p_{0}\equiv p(t_{0})$;
 $\alpha (t)$ is a solution of the nonlinear equation%
\begin{equation}
\frac{\partial _{t}^{2}\alpha (t)}{\alpha (t)}-\omega ^{2}(t)=\frac{4\hslash
^{2}}{\alpha ^{4}(t)}
\end{equation}%
which forms with the linear equation (\ref{20}) an Ermakov pair (see \cite%
{matzkin01} and Refs. therein).\ This means that $\alpha (t)$ can be
expressed in terms of two linearly independent solutions of Eq.(\ref{20}),
the precise choice of a given function $\alpha (t)$ depending on two
arbitrary constants (denoted $I$ and $c$ in \cite{matzkin01}).\ These are
fixed so that initially $\alpha (t_{0})=\alpha _{0}$ and $\alpha ^{\prime
}(t_{0})=0$, as required so that Eq. (\ref{1}) is consistent with Eq.
(\ref{z0}). One then has $q(t)=\sqrt{2I}\alpha (t)\sin \phi (t)$,
where $\phi (t)$ which appears in Eq. (\ref{z0}) is known in the context of
Ermakov systems as the phase function; it is given by $\partial _{t}\phi
(t)\equiv \hslash \alpha ^{-2}(t)$. Note that Ermakov systems have often
been employed in order to study the solutions of the classical time
dependent harmonic oscillator \cite{refs TDHO}. Besides transforming an
ubiquitious nonlinear equation into a linear one, they offer several
advantages: for example by construction $\alpha (t)$ is a positive definite
quadratic form \cite{matzkin01}, ensuring that $\psi (x,t)$ given by Eq. (%
\ref{z0}) is normalizable.

We consider situations in which an initial Gaussian wavefunction (\ref%
{1}) lies far on the negative axis and is launched at $t_{0}$ towards the
right. The wavepacket spreads while travelling to the right (with the spread
controlled by $\alpha(t)$). A detector placed at a point $x_{T}$ far
beyond $q_{0}$ measures the
time-dependent transmission probability by counting the transmitted
particles arriving there up to various instants. At any instant \textit{%
before} the asymptotic value of the reflection probability is attained,
 the time evolving transmission probability in the region $%
x_{T}\leq x<\infty$ is given by
\begin{equation}
T(x_{T,}t) =\int_{x_{T}}^{\infty}\left\vert \psi(x,t)\right\vert ^{2}dx
\label{transprob}
\end{equation}
and then, it follows from Eq. (\ref{z0}) that%
\begin{equation}
T(x_{T,}t)  =\frac{1}{2}\textrm{erfc}\left[ \frac{\sqrt {2m}(x_{T}-q(t))}{\alpha(t)}\right].
\label{transprob2}
\end{equation}

We compute the transmission probabilities for various sets of parameters.
 In order to assess the influence of the appearance of
the time-dependent barrier on the transmitted wavepacket, we set $t_{b}$
in Eq.(\ref{30}) so that the maximum of the free Gaussian reaches $x=0$
when the barrier
strength is the greatest, ie $V(x,t_{b})=-\frac{1}{2}mkx^{2}$ and $%
q_{f}(t_{b})=0$ where $f$ denotes the free case ($k=0$).  Taking
identical initial Gaussians in the free and barrier cases, we can
appropriately choose the initial position and momentum parameters of the
initial wavefunction so that $\psi (x,t)$ and $\psi _{f}(x,t)$ remain almost
identical up to times slightly below $t=t_{b}$. At that point, the rising
barrier perturbs the wavepacket, whereas in the free case the Gaussian keeps
propagating with an average momentum $p_{f}(t)=p_{0}$.
The transmission probabilities which are plotted (as a function
of time) in Fig. 1(a)  in the free case (black curve) and for barriers with
increasing strength (as the colouring goes from blue to red). We denote the
transmitted probability for the free and the barrier-perturbed cases as $%
T_{f}(t)$ and $T_{k}(t)$ respectively. We observe that $T_{k}(t)>T_{f}(t)$
during the time interval $t_{d}<t<t_{c}$ (superarrival). (Here $t_{k}$ is the
instant at which the perturbation starts, $t_{c}$ the instant when the free
and the perturbed curves cross each other, and $t_{d}$ the time from which
the curve corresponding to the perturbed case starts deviating from that in the
free case, so that $t_{c}>t_{d}>t_{k}$).

\paragraph{C. Communication using superarrival. \textemdash}
We have seen above that the transmission probability for the perturbed
barrier exceeds that of the free case in a particular time interval. The
detector during this time interval therefore records more number of
particles than it would have in the free case. At this stage, a
particularly relevant question arises as to when would Bob at the
detector realize that Alice has perturbed the wave packet ? Or, in other words,
how fast does the information about barrier perturbation travel to the
detector ? Bob who records the growth of the transmission probability
becomes aware of the perturbation  (occurring from the instant
$t_{k}$) at the instant $t_{d}$ when the transmission probability starts
deviating from the free case. If Alice (located at the potential barrier)
and Bob are separated by the distance $D$, one can define information velocity $v_{I}$
by
\begin{equation}
v_{I}(k)=\frac{D}{t_{d}-t_{k}}
\end{equation}
We compute $v_{I}(k)$ and plot the function $v_{I}(k)/v_{g}$ (where $v_{g}$
refers to the group velocity of the wave packet in the free case) versus the
strength of the barrier (Fig. 1(b)). Note that information of barrier
perturbation travels from the barrier to the detector with a velocity which
could exceed the group velocity of the wave packet. Fig. 2 displays
superarrivals with parameters chosen so as to enhance the ratio
$v_{I}(k)/v_{g}$ (see Fig. 2(b)).

\begin{figure}[tb]
\includegraphics[height=9cm]{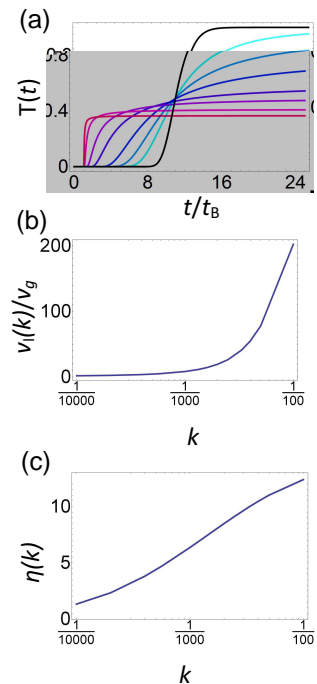}
\caption{Same as Fig. 1 but for a system with parameters $q_0=-10^3, p_0=2, m=1, \alpha^2(t_0)=5, t_B=500, g=1/500$ (atomic units).
The transmission probability $T(t)$ is plotted for the free case
(black curve) and for several time-dependent barriers characterized by
the strengths: $k=1/10000, 1/5000, 1/2500, 1/1000, 1/500, 1/200, 1/100$ (the curve color goes from light blue curve
to red as $k$ increases).}
\label{figT2}
\end{figure}

In order for Alice and Bob to use this information transfer as means of
communication, we define the quantity $\eta $ which determines the
magnitude of superarrival by
\begin{equation}
\eta(k) =\frac{I_{k}-I_{f}}{I_{f}}  \label{8}
\end{equation}%
where $I_{k}$ and $I_{f}$ are defined with respect to the
time interval $\Delta t=t_{c}-t_{d}$ during which superarrival occurs, as
follows:
\begin{equation}
I_{k}=\int_{\Delta t}T_{k}(t)dt; \>\>\>\>
I_{f}=\int_{\Delta t}T_{f}(t)dt  \label{9}
\end{equation}
The magnitude of superarrival $\eta$ is a function of the  barrier
strength $k$. In Figs. 1(c) and 2(c) we plot $\eta$ versus $k$
for a couple of different sets of parameter values.

Now, suppose a
particular functional relation between $k$ and $\eta(k)$ (say, curve 1(c))
is chosen as a key which is shared by Alice and Bob. Alice at the barrier
receives a continuous inflow of particles whose wave function is given by
the initial Gaussian. At first, she does nothing and Bob at the detector
simply records the particle counts, thereby generating the curve $T_f$.
She then introduces the barrier perturbation choosing
random different values for the barrier strength $k$ corresponding to the
different runs of the experiment. For a particular run, Bob
has to decipher this specific value of $k$ chosen by Alice. Bob
monitors the time evolution of $T_k$ through the detector counts, and
by comparing with $T_f$, is able to obtain $\Delta_t$ and compute
$\eta$ using it. Thus, he is able to decipher the value of $k$ chosen
by Alice using the key $\eta(k)$. This whole procedure may be repeated
as many times as required by Alice and Bob in order to exchange any
required (classical) information between them.

The security of this protocol
is checked by using the relation between $k$ and $v_I$ (the curve 1(b)
in this case). The value of $v_I$ can be obtained by Bob by observing $t_d$,
and using pre-determined (with Alice) values of the distance $D$ and the time
$t_k$  at which Alice starts her perturbation. An attempt of
eavesdropping would involve distortion of the wave packet thereby affecting
the correspondence between the barrier strength and the information
velocity. Bob checks whether his deciphered value of $k$ together with the
measured  $v_I$ lie on the curve 1(b). If the pair of values fall outside
the curve, then Bob is able to detect disturbance by the eavesdropper, and
asks Alice to reject that particular run of the experiment.

\begin{figure}[t]
\includegraphics[height=3cm]{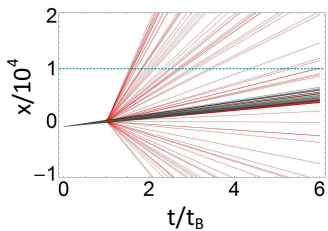}
\caption{Trajectories $q(t)$ for the situation portrayed in Fig. 2 (black
lines: free case; red lines:  time-dependent barrier with $k=1/500$). The
initial conditions for each trajectory are randomly
chosen within the initial distribution  $\rho (x,t_{0})$, centered at $x=-1000$. The barrier potential, maximal at $t=t_B$, highly spreads
the pencil of trajectories, resulting in the diffraction of the wavepacket.}
\label{ClassTr}
\end{figure}

\paragraph{D. Dynamical interpretation. \textemdash}

Superarrivals are produced by the \emph{diffraction} of the wavefunction on
the moving barrier. This
can be seen by resorting to the path integral form of the time evolution
operator $K(x,x^{\prime },t)$.
The Lagrangian with the potential (\ref{30}) is quadratic and
therefore  $K(x,x^{\prime },t)$ is given by%
\begin{equation}
K(x,x^{\prime },t-t_{0})=\sqrt{\frac{i}{2\pi \hslash }\frac{\partial
^{2}S_{cl}}{\partial x\partial x^{\prime }}}\exp \frac{i}{\hslash }%
S_{cl}(x,x^{\prime },t-t_{0}).  \label{36}
\end{equation}%
where $S_{cl}$ is the action for the classical paths going from $x^{\prime }$
to $x$ in time $t-t_{0}$.  Its explicit
expression is cumbersome; it is obtained by employing a method similar to
the one used for
the time-dependent harmonic oscillator (eg Ch.\ 20 of \cite{dittrich
reuter01}), namely by expressing the time integral of the Lagrangian as a
quadratic form in $x$ and $x^{\prime }$. This is precisely done by employing
the Ermakov system decomposition $q(t)=\sqrt{2I}\alpha (t)\sin \phi (t)$.

The important message encapsulated by the propagator expression (\ref{36})
is that every point $x^{\prime }$ of the initial wavefunction $\psi
(x^{\prime },t_{0})$ is carried to the point $\left( x,t\right) $ by a
classical trajectory: a single propagating wavepacket is built on the entire
set of paths
whose initial conditions lie within $\rho (x,t_{0})$ (now regarded as a
configuration space classical distribution).
The corresponding family of trajectories for the superarrival
shown in Fig. 2 is
plotted in Fig. \ref{ClassTr}, along with the free trajectories for the same
case.\ It can be seen that the barrier causes the impinging trajectories to
accelerate or to turn back, depending on their initial positions.
Superarrival is produced by
the paths that are accelerated (relative to the no-barrier situation)
and  therefore arrive at some point $x_{T}$ far to the right of the
barrier in a shorter time than in the free case.

\paragraph{E. Concluding remarks. \textemdash}

To summarize, in this work we have presented a novel quantum effect
in which it is analytically shown that there exists a short time interval
over which  the transmission probability of a Gaussian
wave packet impinging on a transient parabolic potential barrier
exceeds that corresponding to free propagation. Further, we have shown how
this effect can be used for communication across the wave packet.
A simple scheme of information transfer has been outlined which is based
upon the one-to-one correspondence between any particular value of the
barrier strength chosen, and the measured value of the magnitude of
superarrival. Such a scheme of transfer of classical information using the
quantum wave function seems  {\it in-principle} secure, since any
attempt of eavesdropping would involve distortion of the wave packet,
thereby producing a discernible impact on the correspondence between the
barrier strength and the information velocity.
Note that in examples such as
the present one where the incident wave packet gets distorted on striking
the barrier, the concept of signal velocity needs to be defined operationally,
and it can exceed the group velocity of the wave packet \cite{winful}.
Accordingly, we have introduced the notion of information velocity which
measures the speed at which information about the barrier perturbation
propagates across the wave packet.

Finally, a path integral analysis of
superarrivals is invoked to understand how this effect arises from the
diffraction of the wave packet on the barrier.
Further work involving the time-dependent
reflection and transmission of wavepackets from various types of transient
barriers is required for the purpose of experimental tests and applications
of superarrivals.
In particular, one may consider a variant of our example by introducing
two qubits localized on opposite sides of the barrier, and thereby inducing
an entanglement between them by their interaction with the reflected and the
transmitted parts of a wavepacket, in the context of which
superarrivals can be employed in schemes aiming at the speed-up of
entanglement generation.

\textit{Acknowledgements:} ASM and DH acknowledge support from the DST
Project SR/S2/PU-16/2007. DH thanks Centre for Science, Kolkata for useful support.

\end{document}